\documentstyle[aps,epsfig]{revtex}
\begin{document}
\tightenlines
\draft

\newcommand{\be}{\begin{eqnarray}}
\newcommand{\ee}{\end{eqnarray}}
\newcommand{\dia}{\!\!\!\!\!\!\not\,\,\,}
\newcommand{\1}{\'{\i}}

\twocolumn[\hsize\textwidth\columnwidth\hsize\csname
@twocolumnfalse\endcsname

\title{Axially asymmetric fermion scattering off
       electroweak phase transition bubble walls with
       hypermagnetic fields}  
\author{Alejandro Ayala$^1$, Jaime Besprosvany$^2$, Gabriel Pallares$^1$,
       Gabriella Piccinelli$^3$} 
\address{$^1$Instituto de Ciencias Nucleares,
         Universidad Nacional Aut\'onoma de M\'exico,\\
         Apartado Postal 70-543, M\'exico Distrito Federal 04510,
         M\'exico.}
\address{$^2$Instituto de F\1sica,
         Universidad Nacional Aut\'onoma de M\'exico,\\
         Apartado Postal 20-364, M\'exico Distrito Federal 01000,
         M\'exico.}
\address{$^3$Centro Tecnol\'ogico Arag\'on
         Universidad Nacional Aut\'onoma de M\'exico\\
         Av. Rancho Seco S/N, Bosques de Arag\'on, Nezahualc\'oyotl Edo. de
         M\'exico 57130, M\'exico}
\maketitle
\begin{abstract}
We show that in the presence of large scale primordial hypermagnetic  
fields, it is possible to generate an axial asymmetry for
a first order electroweak phase transition. This 
happens during the reflection and transmission of fermions off the
true vacuum bubbles, due to the chiral nature of the fermion
coupling with the background field in the symmetric phase. We derive
and solve the Dirac equation for such fermions and compute the
reflection and transmission coefficients for the case when these
fermions move from the symmetric to the symmetry broken phase. We also
comment on the possible implications of such axial charge segregation
processes for baryon number generation.
\end{abstract}
\vskip2pc]

\section{Introduction}
One of the most challenging problems for particle physics as applied
to cosmology is the explanation of the
observed excess of baryons over antibaryons in the universe. For this
purpose, a theory has to meet the three well-known Sakharov
conditions\cite{Sakharov}, namely: (1) Existence of interactions that
violate baryon number; (2) {\it C} and {\it CP} violation and (3)
departure from thermal equilibrium. The
above conditions are met in the standard model (SM) provided the
electroweak phase transition (EWPT) is of 
first order. This has raised the interesting possibility that the
cosmological phase transition that gave rise to the mass of particles,
which took place at temperatures of order $100$ GeV, could also explain
the generation of baryon number. Consequently, a great deal of effort
has been devoted to explore this possibility~\cite{Trodden}. 

Nowadays, the consensus is that the minimal SM, as such, cannot explain
the observed baryon number. The reason is that the EWPT turns out to
be only too weakly first order which in turn implies that any baryon 
asymmetry generated at the phase transition was erased by the same 
mechanism that produced it, {\it i.e.}, sphaleron
induced processes~\cite{Kajantie}. Moreover, the amount of $CP$ violation 
coming from the CKM matrix alone cannot account by itself for the
observed asymmetry, given that its effect shows up in the coupling of the
Higgs with fermions at a high perturbative order~\cite{Dine}, producing
a baryon to entropy ratio at least ten orders of magnitude
smaller than the observed one.

Nevertheless, it has been recently pointed out that, provided a
source of enough {\it CP} violation exists, the above scenario
could significantly change in the presence of large-scale primordial
magnetic fields~\cite{{Giovannini},{Elmfors},{Giovannini2}} (see however
Ref.~\cite{Skalozub}), which can be responsible for a stronger 
first-order EWPT. This situation is analogous to the case of a type I
superconductor in which the presence of an external magnetic field
modifies the order of the phase transition due to the Meissner
effect. Though the nature of these fields is a subject of current research, 
their existence prior to the EWPT epoch cannot certainly be ruled
out~\cite{reviews}. 

Magnetic fields have been observed in many astrophysical
objects. Estimation of their strengths require independent knowledge
of the local electron density 
and the spatial structure of the field. Both quantities are reasonably
well known for our galaxy, where the average field strength has been
measured to be between $3 - 4 \mu$G; moreover, various spiral galaxies
in our neighborhood present similar magnetic field
strengths~\cite{kron}. At larger scales, only model dependent upper
limits can be established and these are also in the few $\mu$G
range. Magnetic fields at the $\mu$G level have been observed as well in
high-redshift objects. In the intergalactic medium, adopting some
reasonable values for the magnetic coherence length, the upper bound
of $10^{-9}$G has been estimated~\cite{reviews}. The origin of these
fields is nowadays unknown but it is widely believed that, in order to
produce them, two ingredients are needed: a mechanism for creating the
seed fields and a process for amplifying both their amplitude and
their coherence scale.

Generation of the seed field (magnetogenesis) may be either primordial
or be produced during the process of structure formation. In the early
universe, which is the case of interest here, there are a number of
proposed mechanisms that could possibly generate large-scale
primordial fields. Among the best suited are first order phase
transitions~\cite{{Quash},{Baym}}, which 
provide favorable conditions such as charge separation, turbulence and
departure from equilibrium. In particular, bubble wall 
collisions produce phase gradients of a complex order parameter that
act as a source for gauge fields~\cite{Kibble}. 
When interested in larger coherence scales, a plausible scenario is
inflation, where super-horizon scale fields are generated through the
amplification of quantum fluctuations of the gauge 
fields. This process needs however a mechanism for breaking conformal
invariance of the electromagnetic field~\cite{Turner-astro}.

The most promising way to distinguish between primordial and protogalactic
fields is through the search of their imprint on the cosmic microwave
background radiation (CMBR). Temperature anisotropies from COBE
results place an upper bound $B_0\sim 10^{-9}\ $G for homogeneous fields
($B_0$ refers to the intensity that the field would have today under
the assumption of adiabatic decay due to the Hubble
expansion)~\cite{Barrow}. In the case of inhomogeneous fields their
effect must be searched for in the Doppler peaks~\cite{Adams} and in
the polarization of the CMBR~\cite{Kosov}. The future CMBR satellite
missions MAP and PLANCK may reach the required sensitivity for the
detection of these last signals. 

Independently of their origin, primordial fields could have had some
influence on physical processes which occurred in the early universe, like
big-bang nucleosynthesis and electroweak baryogenesis.

Recall that for temperatures above the EWPT, the SU(2)$\times$U(1)$_Y$
symmetry is restored and the propagating, non-screened vector modes that
represent a magnetic field correspond to the U(1)$_Y$
group instead of to the U(1)$_{em}$ group, and are therefore properly
called {\it hypermagnetic} fields. 

In this paper we use a simple model to show that the presence of such
fields also provides a mechanism, working in the same manner as the
existence of additional {\it CP} violation within the SM, to produce
an axial charge segregation during the EWPT. This happens in the
scattering of fermions off the true vacuum bubbles nucleated during
the phase transition and is a consequence of the chiral nature of the
fermion coupling to hypermagnetic fields in the symmetric phase.   

The outline of this work is as follows: In Sect.~\ref{II}, we write
the Dirac equation for the left and right-handed chirality
modes propagating in a background hypermagnetic field during the
EWPT. In Sect.~\ref{III}, we find the solution and discuss its
properties. In Sect.~\ref{IV}, we use this solution to compute
reflection and transmission probabilities. We show that these probabilities 
differ for the two distinct chirality modes. Finally in Sect.~\ref{V},
we conclude by looking out at the possible implications of such axially
asymmetric fermion reflection and transmission.

\section{Dirac equation for fermions moving in a background
         hypermagnetic field}\label{II} 

In a first order phase transition, the conversion from one phase to
another happens through nucleation. The region separating both
phases is called the wall. During the EWPT, the properties of the wall
depend on the effective, finite temperature Higgs potential. Under the
assumption that the wall is thin and that the phase transition happens
when the energy densities of both phases are degenerate, it is
possible to find a one-dimensional analytical solution for the Higgs
field $\phi$ called the {\it kink}. This is given by
\be
   \phi (z) \sim 1 + \tanh (z/\lambda)\, ,
   \label{kink}
\ee
where $z$ is the coordinate along the direction of the phase change and
$\lambda$ is the width of the wall. When scattering is
not affected by diffusion, the problem of fermion reflection and
transmission through the wall can be casted in terms of solving
the Dirac equation with a position dependent fermion mass,
proportional to the Higgs field~\cite{Ayala}. Let us further simplify
the problem by considering the limit when the width of the wall
approaches zero. In this case, the kink solution becomes a step
function, $\Theta (z)$, and consequently, the expression for the
particle's mass becomes 
\be
   m(z)=m_0\Theta (z)\, .
   \label{step}
\ee   
In terms of Eq.~(\ref{step}), we can see that $z\leq 0$ represents
the region outside the bubble, that is the region in the symmetric
phase where particles are massless. Conversely, for $z\geq 0$, the system
is inside the bubble, that is in the broken phase and particles have
acquired a finite mass $m_0$.

In the presence of an external magnetic field, we need to consider
that fermion modes couple differently to the field in the broken and
the symmetric phases. We start the analysis looking at the unbroken
phase.

For $z\leq 0$, the coupling is chiral. Let 
\be
   \Psi_R&=&\frac{1}{2}\left(1 + \gamma_5\right)\Psi\nonumber\\
   \Psi_L&=&\frac{1}{2}\left(1 - \gamma_5\right)\Psi
   \label{chiralmodes}
\ee
represent, as usual, the right and left-handed chirality modes for the
spinor $\Psi$, respectively. Then, the equations of motion for these
modes, as derived from the electroweak interaction Lagrangian, are
\be
   (i\partial\dia\ -\ \frac{y_L}{2}g'A\dia\ )
   \Psi_L - m(z)\Psi_R &=& 0\nonumber\\
   (i\partial\dia\ -\ \frac{y_R}{2}g'A\dia\ )
   \Psi_R - m(z)\Psi_L &=& 0\, ,
   \label{diracsymm}
\ee 
where $y_{R,L}$ are the right and left-handed hypercharges
corresponding to the given fermion, respectively, $g'$ the
$U(1)_Y$ coupling constant and we take $A^\mu=(0,{\mathbf A})$
representing a, not as yet specified, four-vector potential having
non-zero components only for its spatial part, in the rest frame of
the wall. 

The set of Eqs.~(\ref{diracsymm}) can be written as a single equation
for the spinor $\Psi = \Psi_R + \Psi_L$ by adding up the former equations
\be
   \left\{ i\partial\dia\ - A\dia
   \left[\frac{y_R}{4}g'\left(1 + \gamma_5\right)
   +\frac{y_L}{4}g'\left(1 - \gamma_5\right)\right]\right.\nonumber\\
   - m(z)\Big\}\Psi = 0\, .
   \label{diracsingle}
\ee   
Hereafter, we explicitly work in the chiral representation of the
gamma matrices where
\be
   \gamma^0=\!\left(\begin{array}{rr}
   0 & -I \\
   -I & 0 \end{array}\right)\
   \mbox{\boldmath $\gamma$}=\!\left(\begin{array}{rr}
   0 &  \mbox{\boldmath $\sigma$} \\
   \mbox{\boldmath $-\sigma$} & 0 \end{array}\right)\
   \gamma_5=\!\left(\begin{array}{rr}
   I & 0 \\
   0 & -I \end{array}\right)\, .
   \label{gammaschiral}
\ee
Within this representation, we can write Eq.~(\ref{diracsingle}) as
\be
   \Big\{i\partial\dia\ -\ {\mathcal G}A_\mu\gamma^\mu
   -m(z)\Big\}\Psi=0\, ,
   \label{diracsimple}
\ee
where we have introduced the matrix
\be
   {\mathcal G}=\left(\begin{array}{cc}
   \frac{y_L}{2}g'I & 0 \\
   0 & \frac{y_R}{2}g'I \end{array}\right)\, .
   \label{matA}
\ee
We now look at the corresponding equation in the broken-symmetry 
phase. For $z\geq 0$ the coupling of the fermion with the external
field is through the electric charge $e$ and thus, the equation of motion
is simply the Dirac equation describing an electrically
charged fermion in a background magnetic field, namely,
\be
   \Big\{i\partial\dia\ -\ eA_\mu\gamma^\mu
   -m(z)\Big\}\Psi=0\, .
   \label{diracsimplezg0}
\ee
In the following section, we explicitly construct the solutions to
Eqs.~(\ref{diracsimple}) and ~(\ref{diracsimplezg0}) with a constant
magnetic field, requiring that these match at the interface $z=0$.
    
\section{Solving the Dirac equation}\label{III}

Let us first find the solution to Eq.~(\ref{diracsimple}), namely, for
fermions moving in the symmetric phase, $z\leq 0$. For this purpose,
we look for a solution of the form
\be
   \Psi = \Big\{i\partial\dia\ -\ A_\mu\gamma^\mu{\mathcal G}
   +m(z)\Big\}\Phi\, .
   \label{form}
\ee
Inserting this expression into Eq.~(\ref{diracsimple}), we obtain
\be
   \Big\{-&\partial^2& - i{\mathcal G}\partial^\mu A_\mu - \frac{1}{2}
   \sigma^{\mu\nu}{\mathcal G}F_{\mu\nu} 
   -\nonumber\\ &2i&{\mathcal G}A_\mu\partial^\mu + 
   {\mathcal G}^2A_\mu A^\mu + 
   i\gamma^\mu \partial_\mu m(z)\ \Big\}\ 
   \Phi =0\, ,
   \label{insert}
\ee   
where, as usual,
\be
   \sigma^{\mu\nu}&=&\frac{i}{2}\left[\gamma^\mu,
   \gamma^\nu\right]\nonumber\\
   F_{\mu\nu}&=&\partial_\mu A_\nu - \partial_\nu A_\mu\, .
   \label{usual}
\ee
For definiteness, let us consider a constant magnetic field
${\mathbf B}=B\hat{z}$ pointing along the $\hat{z}$ direction. In this
case, the vector potential ${\mathbf A}$ can only have components
perpendicular to $\hat{z}$ and the solution to Eq.~(\ref{insert})
factorizes as~\cite{Cea} 
\be
   \Phi (t,\mbox{\boldmath $x$})=\zeta (x,y)\Phi (t,z)\, .
   \label{facto}
\ee
We concentrate on the solution describing the motion of fermions
perpendicular to the wall, {\it i.e.}, along the $\hat{z}$
axis and, furthermore, look for stationary states, namely
\be
   \Phi (t,z)=e^{-iEt}\Phi (z)\, .
   \label{stationary}
\ee
Therefore, working in the Lorentz gauge, $\partial^\mu A_\mu =0$,
Eq.~(\ref{insert}) becomes 
\be
   \Big\{\frac{d^2}{dz^2} 
   + i\gamma^3\frac{dm(z)}{dz} + E^2 
   + iB{\mathcal G}\gamma^1\gamma^2 \Big\}
   \Phi (z) = 0\, .
   \label{ecz}
\ee   
Notice that Eqs.~(\ref{insert}) and~(\ref{ecz}) have the appropriate
limit when $y_R=y_L=e$, corresponding to the description of fermions
coupled with their electric charge to a background magnetic field~\cite{Cea}.

We now expand $\Phi (z)$ in terms of the eigen-spinors $u^s_\pm$
$(s=1,2)$ of $\gamma^3$,
\be
   u^1_\pm = \left(\begin{array}{r}
                1 \\ 0 \\ \pm i \\ 0
                \end{array}
              \right)\ \ \
   u^2_\pm = \left(\begin{array}{r}
                0 \\ 1 \\ 0 \\ \mp i
                \end{array}
              \right)\, .
   \label{spinors}
\ee
These spinors have the properties
\be
   \gamma^3u^{1,2}_\pm&=&\pm iu^{1,2}_\pm\,\nonumber\\
   \gamma^0u^1_\pm&=&\mp iu^1_\mp\,\nonumber\\
   \gamma^0u^2_\pm&=&\pm iu^2_\mp\,\nonumber\\
   \gamma^1\gamma^2u^1\pm&=&-iu^1_\pm\,\nonumber\\
   \gamma^1\gamma^2u^2\pm&=&+iu^2_\pm\,\nonumber\\
   \gamma_5u^{1,2}_\pm&=&u^{1,2}_\mp\, .
   \label{gammaspinor}
\ee
Writing
\be
  \Phi (z) = \phi^1_+(z)u^1_+ + \phi^1_-(z)u^1_-
           + \phi^2_+(z)u^2_+ + \phi^2_-(z)u^2_-
  \label{expand}
\ee 
and inserting this expression into Eq.~(\ref{ecz}), we obtain
\be
   \left[\frac{d^2}{dz^2} + E^2 +
   g'\frac{(y_L+y_R)}{4}B\right]\phi^1_+(z) &+&\nonumber\\
   g'\frac{(y_L-y_R)}{4}B\phi^1_-(z) &=&\nonumber\\ 
   m_0\delta (z)\phi^1_+(z)\nonumber\\
   \left[\frac{d^2}{dz^2} + E^2 +
   g'\frac{(y_L+y_R)}{4}B\right]\phi^1_-(z) &+&\nonumber\\
   g'\frac{(y_L-y_R)}{4}B\phi^1_+(z) &=&\nonumber\\ 
   -m_0\delta (z)\phi^1_-(z)
   \label{set1}
\ee
and
\be
   \left[\frac{d^2}{dz^2} + E^2 -
   g'\frac{(y_L+y_R)}{4}B\right]\phi^2_+(z) &-&\nonumber\\
   g'\frac{(y_L-y_R)}{4}B\phi^2_-(z) &=&\nonumber\\ 
   m_0\delta (z)\phi^2_+(z)\nonumber\\
   \left[\frac{d^2}{dz^2} + E^2 -
   g'\frac{(y_L+y_R)}{4}B\right]\phi^2_-(z) &-&\nonumber\\
   g'\frac{(y_L-y_R)}{4}B\phi^2_+(z) &=&\nonumber\\
   -m_0\delta (z)\phi^2_-(z)\, .
   \label{set2}
\ee
Equations~(\ref{set1}) and~(\ref{set2}), represent, each, a set of two
coupled second-order differential equations. The second set is
obtained from the first one by changing $B$ to $-B$. Consequently,
Eqs.~(\ref{set1}) and the corresponding functions and
spinors with $s=1$ describe the motion of the spin components parallel
to to magnetic field whereas Eqs.~(\ref{set2}) and the functions and
spinors with $s=2$, describe the motion of
the spin components antiparallel to the magnetic field. Notice that in
the limit when $y_R=y_L=e$, each set of equations decouple as is the
case when describing the interaction of fermions with the magnetic
field through their electric charge.

Let us focus on the set of Eqs.~(\ref{set1}), since,
as we have pointed out, the solutions to Eqs.~(\ref{set2}) are
obtained from those to Eqs.~(\ref{set1}) by changing $B$ to $-B$.

To solve Eqs.~(\ref{set1}), we look for the scattering states
appropriate to describe the motion of fermions in the symmetric
phase. For our purposes, these are fermions incident towards and
reflected from the wall. There are two types of such solutions; those
coupled with $y_L$ and those coupled with $y_R$. For an incident wave
coupled with $y_L$ ($y_R$), the fact that the differential
Eqs.~(\ref{set1}) mix up the solutions means that the reflected wave
will also include a component coupled with $y_R$ ($y_L$). Let us classify the
solutions according to the type of wave that is incident towards the
wall. For an incident wave coupled with $y_L$, which we call
type $(a)$, the solutions $\phi^{1(a)}_\pm (z)$ are
\be
   \phi^{1(a)}_\pm (z) &=& e^{i\alpha^L_1z} - 
   \frac{m_0^2}{4\alpha^L_1\alpha^R_1 + m_0^2}e^{i\alpha^L_1|z|}\nonumber\\
   &\mp&
   \frac{2im_0\alpha^L_1}{4\alpha^L_1\alpha^R_1 + m_0^2}e^{i\alpha^R_1|z|}\, ,
   \label{typea}
\ee
whereas, for an incident wave coupled with $y_R$, which we call
type $(b)$, the solutions $\phi^{1(b)}_\pm (z)$ are
\be
   \phi^{1(b)}_\pm (z) &=& \pm e^{i\alpha^R_1z} - 
   \frac{2im_0\alpha^R_1}{4\alpha^L_1\alpha^R_1 + m_0^2}
   e^{i\alpha^L_1|z|}\nonumber\\
   &\mp&
   \frac{m_0^2}{4\alpha^L_1\alpha^R_1 + m_0^2}e^{i\alpha^R_1|z|}\, ,
   \label{typeb}
\ee
where we use the notation
\be
   \alpha^{R,L}_1=\sqrt{E^2 + \frac{y_{R,L}g'}{2}B}\, .
   \label{alphaRL1}
\ee
It is a straightforward exercise to verify that the functions
$\phi^{1(a,b)}_\pm (z)$ given by Eqs.~(\ref{typea}) and~(\ref{typeb})
indeed satisfy the system of Eqs.~(\ref{set1}).

The corresponding fermion wave functions are given in terms of
Eq.~(\ref{form}). Taking $E>0$ and in the approximation where we look
only at the part of the wave function that describes motion
perpendicular to the wall, we obtain, for solutions type $(a)$ 
\be
   \Psi_{\mbox {\small inc}}^{(a)}(z)&=&-i(\alpha^L_1-E)(u^1_+-u^1_-)
   e^{i\alpha_1^Lz}\nonumber\\
   &-&i (\alpha^L_2+E)(u^2_+-u^2_-)e^{i\alpha_2^Lz}\nonumber\\
   \Psi_{\mbox{\small ref}}^{(a)}(z)&=&
   \frac{-im_0}{4\alpha_1^L\alpha_1^R+m_0^2}
   \Big\{m_0(\alpha^L_1+E)(u^1_+-u^1_-)e^{-i\alpha_1^Lz}\nonumber\\ 
   &+&2i\alpha_1^L(\alpha^R_1-E)(u^1_++u^1_-)e^{-i\alpha_1^Rz}\Big\}
   \nonumber\\
   &-&\frac{im_0}{4\alpha_2^L\alpha_2^R+m_0^2}
   \Big\{m_0(\alpha^L_2-E)(u^2_+-u^2_-)e^{-i\alpha_2^Lz}\nonumber\\ 
   &+&2i\alpha_2^L(\alpha^R_2+E)(u^2_++u^2_-)e^{-i\alpha_2^Rz}\Big\}\, ,
   \label{increfa} 
\ee
whereas for solutions type $(b)$
\be
   \Psi_{\mbox {\small inc}}^{(b)}(z)&=&-i(\alpha^R_1+E)(u^1_++u^1_-)
   e^{i\alpha_1^Rz}\nonumber\\
   &-&i (\alpha^R_2-E)(u^2_++u^2_-)e^{i\alpha_2^Rz}\nonumber\\
   \Psi_{\mbox{\small ref}}^{(b)}(z)&=&
   \frac{-im_0}{4\alpha_1^L\alpha_1^R+m_0^2}
   \Big\{m_0(\alpha^R_1-E)(u^1_++u^1_-)e^{-i\alpha_1^Rz}\nonumber\\ 
   &+&2i\alpha_1^R(\alpha^L_1+E)(u^1_+-u^1_-)e^{-i\alpha_1^Lz}\Big\}
   \nonumber\\
   &-&\frac{im_0}{4\alpha_2^L\alpha_2^R+m_0^2}
   \Big\{m_0(\alpha^R_2+E)(u^2_++u^2_-)e^{-i\alpha_2^Rz}\nonumber\\ 
   &+&2i\alpha_2^R(\alpha^L_2-E)(u^2_+-u^2_-)e^{-i\alpha_2^Lz}\Big\}\, ,
   \label{increfb} 
\ee
where, in analogy with Eq.~(\ref{alphaRL1}), we define
\be
   \alpha^{R,L}_2=\sqrt{E^2 - \frac{y_{R,L}g'}{2}B}\, .
   \label{alphaRL2}
\ee
We now turn to finding the solution to Eq.~(\ref{diracsimplezg0}),
namely, for fermions moving in the broken phase, $z\geq
0$. This time, we look for a solution of the form
\be
   \Psi = \Big\{i\partial\dia\ -e\ A_\mu\gamma^\mu
   +m(z)\Big\}\Phi\, .
   \label{form2}
\ee
By a procedure similar to that leading to Eqs.~(\ref{set1})
and~(\ref{set2}), the corresponding equations for the functions
$\phi^{1,2}_\pm (z)$ in this region become
\be
   \left[\frac{d^2}{dz^2} + E^2 - m_0^2 +
   eB\right]\phi^1_\pm (z) &=& 
   \pm m_0\delta (z)\phi^1_\pm (z)\nonumber\\
   \left[\frac{d^2}{dz^2} + E^2 - m_0^2 -
   eB\right]\phi^2_\pm (z) &=& 
   \pm m_0\delta (z)\phi^2_\pm (z)\, .
\label{setzg0}
\ee
As expected, when the coupling of the fermion with the external
magnetic field is through its electric charge, the equations
describing the behavior of the functions $\phi^{1,2}_\pm (z)$
decouple. For our purposes, we look for the scattering states
appropriate for the description of transmitted waves. These are
\be
   \phi^{1,2}_\pm (z)=e^{i\alpha_{1,2} z}\mp 
   \frac{im_0}{2\alpha_{1,2} \pm im_0}e^{i\alpha_{1,2} |z|}\, ,
   \label{solzg0}
\ee
where we use the notation
\be
   \alpha_1&=&\sqrt{E^2-m_0^2+eB}\nonumber\\
   \alpha_2&=&\sqrt{E^2-m_0^2-eB}\, .
   \label{alpha12}
\ee
It is also a straightforward exercise to verify that
Eq.~(\ref{solzg0}) indeed satisfies the set of
Eqs.~(\ref{setzg0}). The fermion wave function is obtained from
Eq.~(\ref{form2}). Also, for $E>0$ and in the approximation where we
look only at the part describing the motion of 
particles along $\hat{z}$ and furthermore, imposing continuity of the
fermion wave function at $z=0$, we obtain for solutions type $(a)$
\be
   \Psi_{\mbox {\small tra}}^{(a)}(z)&=&
   \frac{2\alpha^L_1}{4\alpha_1^L\alpha_1^R+m_0^2}
   \Big\{m_0(\alpha^R_1-E)(u^1_++u^1_-)\nonumber\\ 
   &-&i\ [2\alpha_1^R(\alpha_1^L-E)+m_0^2](u^1_+-u^1_-)\Big\}e^{i\alpha_1z}
   \nonumber\\
   &+&\frac{2\alpha^L_2}{4\alpha_2^L\alpha_2^R+m_0^2}
   \Big\{m_0(\alpha^R_2+E)(u^2_++u^2_-)\nonumber\\ 
   &-&i\ [2\alpha_2^R(\alpha_2^L+E)+m_0^2](u^2_+-u^2_-)\Big\}e^{i\alpha_2z},
   \label{transa}
\ee
and for solutions type $(b)$
\be
   \Psi_{\mbox {\small tra}}^{(b)}(z)&=&
   \frac{2\alpha^R_1}{4\alpha_1^L\alpha_1^R+m_0^2}
   \Big\{m_0(\alpha^L_1+E)(u^1_+-u^1_-)\nonumber\\ 
   &-&i\ [2\alpha_1^L(\alpha_1^R+E)+m_0^2](u^1_++u^1_-)\Big\}e^{i\alpha_1z}
   \nonumber\\
   &+&\frac{2\alpha^R_2}{4\alpha_2^L\alpha_2^R+m_0^2}
   \Big\{m_0(\alpha^L_2-E)(u^2_+-u^2_-)\nonumber\\ 
   &-&i\ [2\alpha_2^L(\alpha_2^R-E)+m_0^2](u^2_++u^2_-)\Big\}e^{i\alpha_2z}.
   \label{transb}
\ee
Recall that in the absence of the hypermagnetic field, the eigenvalues
of the chirality and the helicity operators, ($\chi$ and $h$,
respectively) are the same. The presence of the external field lifts such 
degeneracy and the eigenstates of chirality no longer have a
definite helicity. Nevertheless, it is easy to check that for field
strengths $eB$ smaller than $m_0^2$, the component with $h$ that would
correspond to a given $\chi$  in the absence of the external
field, dominates over the rest of the components. For $E>0$, this
means that, to a good approximation, left (right)-handed particles 
are transmitted as such (both in chirality and helicity) but become
right (left)-handed (both in chirality and helicity) upon
reflection. In these cases and to a good approximation, 
the quantum number conserved during scattering off the wall is the
ratio $\chi/h=1$. It can also be shown~\cite{Ayala2} that to a good
approximation, for $E<0$, the corresponding conserved quantum number
is $\chi/h=-1$.    

\section{Reflection and transmission probabilities}\label{IV}

The fact that the amplitudes in Eqs.~(\ref{increfb}) and~(\ref{transb})
are not the same as those in Eqs.~(\ref{increfa}) 
and~(\ref{transa}), means that an axial asymmetry is built during
the scattering of fermions off the wall. To quantify the asymmetry, we
need to compute the corresponding reflection and transmission
coefficients. These are built from the reflected, transmitted and
incident currents of each type. Recall that for a given spinor wave
function $\Psi$, the current normal to the wall is given by
\be
   J=\Psi^\dagger\gamma^0\gamma^3\Psi\, .
   \label{current}
\ee
As can be seen from Eqs.~(\ref{increfb}),~(\ref{transb}) 
and~(\ref{increfa}),~(\ref{transa}), an incident wave with a given chirality
(left-handed for waves type $(a)$, right-handed for waves type $(b)$,
contains, upon reflection and transmission, both kinds of chirality
modes. For waves type $(a)$, the corresponding currents are
\be
   J_{\mbox{\small inc}}^{(a)}&=&4\Big\{(\alpha_2^L+E)^2 -
   (\alpha_1^L-E)^2\Big\}\nonumber\\
   J_{\mbox{\small ref}}^{(a)}&=&J_{\mbox{\small ref}}^{(a)R}
   +J_{\mbox{\small ref}}^{(a)L}
   \nonumber
\ee
where
\be
   J_{\mbox{\small ref}}^{(a)R}&=&-4m_0^2\Big\{\left(\frac{2\alpha^L_2}
   {4\alpha^L_2\alpha^R_2+m_0^2}\right)^2(\alpha^R_2+E)^2\nonumber\\
   &-&\left(\frac{2\alpha^L_1}{4\alpha^L_1\alpha^R_1+m_0^2}\right)^2
   (\alpha^R_1-E)^2\Big\}\nonumber\\
   J_{\mbox{\small ref}}^{(a)L}&=&-4m_0^2\Big\{
   \left(\frac{m_0}{4\alpha^L_1\alpha^R_1+m_0^2}\right)^2
   (\alpha^L_1+E)^2\nonumber\\
   &-&\left(\frac{m_0}{4\alpha^L_2\alpha^R_2+m_0^2}\right)^2
   (\alpha^L_2-E)^2\Big\}
   \nonumber
\ee
and
\be
   J_{\mbox{\small tra}}^{(a)}&=&J_{\mbox{\small tra}}^{(a)R}
   +J_{\mbox{\small tra}}^{(a)L}
   \nonumber
\ee
where
\be
   J_{\mbox{\small tra}}^{(a)R}&=&16\Big\{
   \left(\frac{m_0\alpha^L_1(\alpha^R_1-E)}
   {4\alpha^L_1\alpha^R_1+m_0^2}\right)^2\nonumber\\
   &-&\left(\frac{m_0\alpha^L_2(\alpha^R_2+E)}
   {4\alpha^L_2\alpha^R_2+m_0^2}\right)^2\Big\}\nonumber\\
   J_{\mbox{\small tra}}^{(a)L}&=&16\Big\{\left(
   \frac{2\alpha^L_2\alpha^R_2(\alpha^L_2+E)+m_0^2\alpha^L_2}
   {4\alpha^L_2\alpha^R_2+m_0^2}\right)^2\nonumber\\
   &-&\left(\frac{2\alpha^L_1\alpha^R_1(\alpha^L_1-E)+m_0^2\alpha^L_1}
   {4\alpha^L_1\alpha^R_1+m_0^2}\right)^2\Big\}\, ,
   \label{ja}
\ee
whereas for waves of type $(b)$, the corresponding currents are
\be
   J_{\mbox{\small inc}}^{(b)}&=&4\Big\{(\alpha_1^R+E)^2 -
   (\alpha_2^R-E)^2\Big\}\nonumber\\
   J_{\mbox{\small ref}}^{(b)}&=&J_{\mbox{\small ref}}^{(b)R}+
   J_{\mbox{\small ref}}^{(b)L}
   \nonumber
\ee
where
\be
   J_{\mbox{\small ref}}^{(b)R}&=&-4m_0^2\Big\{
   \left(\frac{m_0}{4\alpha^L_2\alpha^R_2+m_0^2}\right)^2
   (\alpha^R_2+E)^2\nonumber\\
   &-&\left(\frac{m_0}{4\alpha^L_1\alpha^R_1+m_0^2}\right)^2
   (\alpha^R_1-E)^2\Big\}\nonumber\\
   J_{\mbox{\small ref}}^{(b)L}&=&-4m_0^2\Big\{
   \left(\frac{2\alpha^R_1}
   {4\alpha^L_1\alpha^R_1+m_0^2}\right)^2(\alpha^L_1+E)^2\nonumber\\
   &-&\left(\frac{2\alpha^R_2}
   {4\alpha^L_2\alpha^R_2+m_0^2}\right)^2(\alpha^L_2-E)^2
   \Big\}
   \nonumber
\ee
and
\be
   J_{\mbox{\small tra}}^{(b)}&=&J_{\mbox{\small tra}}^{(b)R}+
   J_{\mbox{\small tra}}^{(b)L}
   \nonumber
\ee
where
\be
   J_{\mbox{\small tra}}^{(b)R}&=&16\Big\{
   \left(\frac{2\alpha^L_1\alpha^R_1(\alpha^R_1+E)+m_0^2\alpha^R_1}
   {4\alpha^L_1\alpha^R_1+m_0^2}\right)^2\nonumber\\
   &-&\left(\frac{2\alpha^L_2\alpha^R_2(\alpha^R_2-E)+m_0^2\alpha^R_2}
   {4\alpha^L_2\alpha^R_2+m_0^2}\right)^2\Big\}\nonumber\\
   J_{\mbox{\small tra}}^{(b)L}&=&16\Big\{
   \left(\frac{m_0\alpha^R_2(\alpha^L_2-E)}
   {4\alpha^L_2\alpha^R_2+m_0^2}\right)^2\nonumber\\
   &-&\left(\frac{m_0\alpha^R_1(\alpha^L_1+E)}
   {4\alpha^L_1\alpha^R_1+m_0^2}\right)^2\Big\}\, .
   \label{jb}
\ee
The reflection and transmission coefficients are given as the ratios of
the reflected and transmitted currents, to the incident one,
respectively, projected along a unit vector normal to the
wall,
\be
   R_{L\rightarrow L}&=&
   -J_{\mbox{\small ref}}^{(a)L}/
   J_{\mbox{\small inc}}^{(a)}\nonumber\\
   R_{L\rightarrow R}&=&
   -J_{\mbox{\small ref}}^{(a)R}/
   J_{\mbox{\small inc}}^{(a)}\nonumber\\
   T_{L\rightarrow L}&=&
   J_{\mbox{\small tra}}^{(a)L}/
   J_{\mbox{\small inc}}^{(a)}\nonumber\\
   T_{L\rightarrow R}&=&
   J_{\mbox{\small tra}}^{(a)R}/
   J_{\mbox{\small inc}}^{(a)}\, ,
   \label{RFa}
\ee
Equations~(\ref{RFa}) represent the probabilities that a left-handed
incident particle bounces off the wall as a left or a right-handed
particle or is transmitted through the wall as a left or a
right-handed particle, respectively. The corresponding probabilities
for the axially conjugate processes are 
\be
   R_{R\rightarrow L}&=&
   -J_{\mbox{\small ref}}^{(b)L}/
   J_{\mbox{\small inc}}^{(b)}\nonumber\\
   R_{R\rightarrow R}&=&
   -J_{\mbox{\small ref}}^{(b)R}/
   J_{\mbox{\small inc}}^{(b)}\nonumber\\
   T_{R\rightarrow L}&=&
   J_{\mbox{\small tra}}^{(b)L}/
   J_{\mbox{\small inc}}^{(b)}\nonumber\\
   T_{R\rightarrow R}&=&
   J_{\mbox{\small tra}}^{(b)R}/
   J_{\mbox{\small inc}}^{(b)}\, .
   \label{RFb}
\ee 
Therefore, the probabilities for finding a left or a right-handed
particle in the symmetric phase after reflection, $PR_L$, $PR_R$ are
given, respectively by
\be
   PR_L&=&R_{L\rightarrow L}+R_{R\rightarrow L}\nonumber\\
   PR_R&=&R_{L\rightarrow R}+R_{R\rightarrow R}\, ,
   \label{PR}
\ee
whereas the probabilities for finding a left or a right-handed
particle in the symmetry broken phase after transmission, $PT_L$, $PT_R$ are
given, respectively by
\be
   PT_L&=&T_{L\rightarrow L}+T_{R\rightarrow L}\nonumber\\
   PT_R&=&T_{L\rightarrow R}+T_{R\rightarrow R}\, .
   \label{PT}
\ee
Figure~1 shows the probabilities $PR_L$ and $PR_R$ as a function of the
magnetic field parametrized as $B=bT^2$ for a temperature $T=100$ GeV,
a fixed $E=184$ GeV and for a fermion taken as the top quark with a
mass $m_0=175$ GeV, $y_R=4/3$, $y_L=1/3$ and for a value of
$g'=0.344$, as appropriate for the EWPT epoch. Notice that when
$b\rightarrow 0$, these probabilities approach each other and that the
difference grows with increasing field strength. Also, in order to
be able to safely neglect the contribution from the negative energy
solutions, we are bound to consider not too large values of the
parameter $b$. For the purposes of this work, we take a maximum value
of $b=1$ which for the values of $T$ and $m_0$ considered, amounts
for a maximum fraction of the magnetic energy to the particle's rest mass of
order $\sqrt{eB}/m_0\sim 0.3$.

\begin{figure}[t] 
\centerline{\epsfig{file=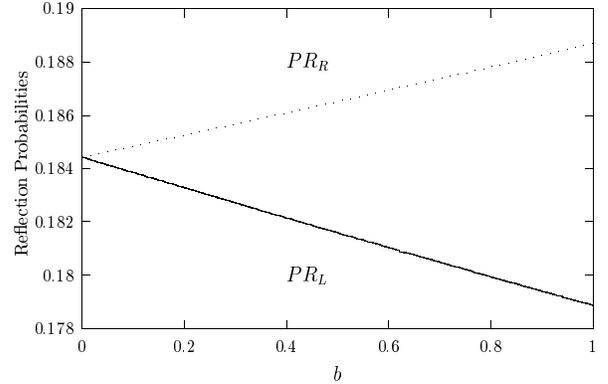,height=2.0in,width=3in}
}
\vspace{0.5cm}
\caption{Probabilities $PR_L$ and $PR_R$ as a function of the magnetic
field parametrized as $B=bT^2$ for $T=100$ GeV, $E=184$ GeV and a top
quark with a mass $m_0=175$ GeV, $y_R=4/3$, $y_L=1/3$. The value for
the $U(1)_Y$ coupling constant is taken as $g'=0.344$, corresponding
to the EWPT epoch.} 
\end{figure}

Figure~2 shows the reflection and transmission probabilities as a
function of the particle's energy $E$. Figure~2a shows the
probabilities $PR_L$ and $PT_L$ and Fig.~2b the probabilities $PR_R$
and $PT_R$ for $b=1$. As before, the parameters considered correspond
to a top quark. Since the solutions in Eqs.~(\ref{transa})
and~(\ref{transb}) are computed assuming that the transmitted waves
are not exponentially damped, the energy has to be taken such that the
parameters $\alpha_{1,2}$ in Eqs.~(\ref{alpha12}) are real, which in
turn implies that $E\geq\sqrt{m_0^2+eB}$. It can be numerically checked that 
$PR_L + PT_L = PR_R + PT_R = 1$ to within a maximum deviation of one
part in one thousand. The fact that these probabilities add up to one
is equivalent to current conservation 
\be
   J_{\mbox{\small tra}}^{(i)}-J_{\mbox{\small ref}}^{(i)}=
   J_{\mbox{\small inc}}^{(i)}\, ,
   \label{adduptoone}
\ee
($i=a,b$), as a consequence of the equality of the currents
\be
   J_{\mbox{\small tra}}^{(a)R}&=&J_{\mbox{\small ref}}^{(a)R}\nonumber\\
   J_{\mbox{\small tra}}^{(b)L}&=&J_{\mbox{\small ref}}^{(b)L}\, ,
   \label{equal}
\ee
as can be checked from the sets of Eqs.~(\ref{ja}) and~(\ref{jb}).

\section{Conclusions and outlook}\label{V}

In this paper we have derived and solved the Dirac equation for
fermions scattering off a first order EWPT bubble wall in the presence of a
magnetic field directed along the fermion direction of motion. In the
symmetric phase, the fermions couple chirally to the magnetic field,
which receives the name of {\it hypermagnetic}, given that it
belongs to the $U(1)_Y$ group. We have shown that the chiral nature of
this coupling implies that it is possible to build an axial
asymmetry during the scattering of fermions off the
wall. We have computed reflection and transmission coefficients
showing explicitly that they differ for left and right-handed incident
particles from the symmetric phase. 

Recall that under the very general assumptions of CPT invariance,
together with conservation of unitarity, which are satisfied in the
present analysis, the total axial asymmetry (which includes
contributions both from particles and antiparticles) is quantified in
terms of the particle (axial) asymmetry. Let $\rho_i$ represent the
number density for species $i$. The net densities in left-handed
and right-handed axial charges are obtained by taking the differences
$\rho_L-\rho_{\bar{L}}$ and $\rho_R-\rho_{\bar{R}}$, respectively. It
is straightforward to show~\cite{Nelson} that CPT invariance and
unitarity imply that the above net densities are given by
\be
   \rho_L-\rho_{\bar{L}}&=&(f^s-f^b)(PR_L - PR_R)\nonumber\\
   \rho_R-\rho_{\bar{R}}&=&(f^s-f^b)(PR_R - PR_L)\, ,
   \label{net} 
\ee  
where $f^s$ and $f^b$ are the statistical distributions for
particles or antiparticles (since the chemical potentials are assumed
to be zero or small compared to the temperature, these distributions
are the same for particles or antiparticles) in the symmetric and the
symmetry-broken phases, respectively. From Eq.~(\ref{net}), the
asymmetry in the axial charge density is finally given by
\begin{equation}
   (\rho_L-\rho_{\bar{L}}) - (\rho_R-\rho_{\bar{R}})=
   2(f^s-f^b)(PR_L - PR_R)\, .
   \label{final}
\end{equation}

This asymmetry in the axial charge, built on either side of the
wall, is dissociated from non-conserving baryon number 
processes and can subsequently be converted to baryon number in the unbroken
phase where sphaleron induced transitions are taking place with a large
rate. This mechanism receives the name of 
{\it non-local baryogenesis}~\cite{{Nelson},{Dine2},{Cohen},{Joyce}}
and, in the absence of the external field, it can only be realized in
extensions of the SM where a source of {\it CP} violation
is introduced {\it ad hoc} into a complex, space-dependent phase of
the Higgs field during the development of the EWPT~\cite{Torrente}.

\begin{figure}[t!] 
\centerline{\epsfig{file=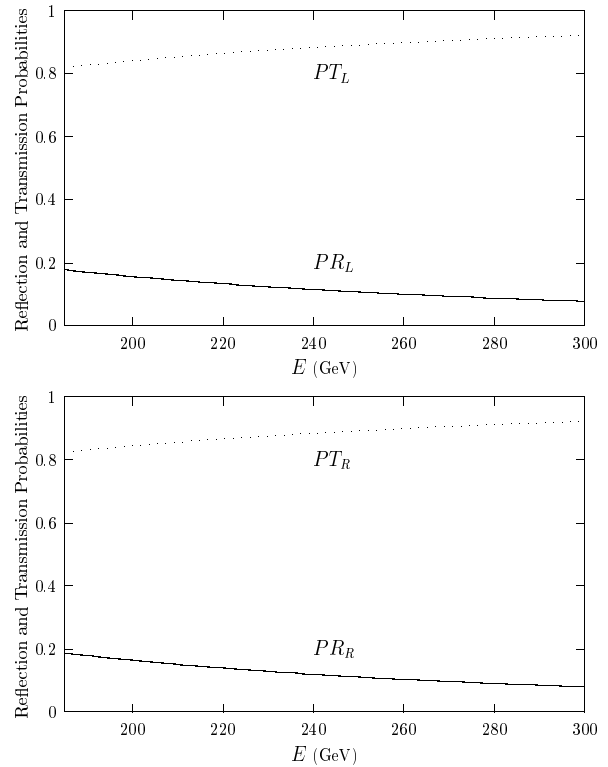,height=4.0in,width=3in}
}
\vspace{0.5cm}
\caption{Reflection and transmission probabilities as a
function of the particle's energy $E$. Figure 2a (upper panel) shows the
probabilities $PR_L$ and $PT_L$. Figure~2b
(lower panel) shows the probabilities $PR_R$ and 
$PT_R$. In both cases, the strength of the magnetic
field is taken with $b=1$ and $T=100$ GeV. Also $m_0=175$ GeV,
$y_R=4/3$, $y_L=1/3$, as corresponds to a top quark.}
\end{figure}

Due to the sphaleron dipole moment, another consequence of the
existence of an external magnetic field is the lowering of the barrier
between topologically inequivalent vacuua~\cite{Comelli}. This effect
acts in such a way that any baryon asymmetry generated by the
building of an axial charge during the 
asymmetric reflection of fermions into the unbroken phase, in the
presence of a magnetic field, stands little chance of surviving in the
broken phase. Nonetheless, if such primordial fields indeed
existed during the EWPT epoch and the phase transition was
first order, as is the case, for instance, in minimal extensions of the 
SM, the mechanism advocated in this work has to be considered as
acting in the same manner as a source of {\it CP} violation that can
have important consequences for the generation of a baryon
number. These matters will be the subject of an upcoming work~\cite{Ayala2}.

\section*{Acknowledgments}

Support for this work has been received in part by CONACyT-Mexico
under grant number 32279-E and by DGAPA-UNAM under grant number
IN118600.

\end{document}